\documentclass{article} 
\usepackage{graphicx,amsmath}
\usepackage{amsmath,amsfonts,latexsym,amssymb}
\usepackage{verbatim,amsthm,curves,graphics}
\usepackage{mathrsfs}

\def\ben{\begin{equation}}
\def\een{\end{equation}}
\def\bena{\begin{eqnarray}}
\def\eena{\end{eqnarray}}

\def\f(#1/#2){\frac{#1}{#2}} 
\def\Frac(#1/#2){\left(\frac{#1}{#2}\right)} 
\def\chris(#1-#2-#3){{\mit \Gamma}^{#1}{}_{{#2}{#3}} }
\def\tilchris(#1-#2-#3){\tilde{{\mit \Gamma}}^{#1}{}_{{#2}{#3}}}
\def\hatchris(#1-#2-#3){\hat{{\mit \Gamma}}^{#1}{}_{{#2}{#3}}}  

\def\tilg{\tilde{g}} 
\def\tilR{\tilde{R}}

\newcommand{\non}{\nonumber}
\theoremstyle{definition}

\newcommand{\I}{{\mathscr I}}
\renewcommand{\S}{{\mathscr S}}

\newcommand{\tg}{{\tilde g}}

\begin{document}

\title{Asymptotic flatness at null infinity in higher dimensional gravity}

\author{\it Stefan Hollands$^1$ and Akihiro Ishibashi$^2$\\
\rm $^1$Enrico Fermi Institute, Department of Physics, \\
University of Chicago \\
5640 Ellis Ave., Chicago IL 60637 (USA) \\
E-mail: stefan@bert.uchicago.edu\\
$^2$Department of Applied Mathematics and Theoretical Physics, \\
University of Cambridge, \\ 
Wilberforce Road, Cambridge CB3 0WA (UK) \\ 
E-mail: A.Ishibashi@damtp.cam.ac.uk}

\maketitle

\abstract{ 
We give a geometrical definition of the asymptotic flatness 
at null infinity in spacetimes of even dimension $d$ greater 
than $4$ within the framework of conformal infinity. 
Our definition is shown to be stable against perturbations to linear order.  
We also show that our definition is stringent enough to allow 
one to define the total energy of the system viewed from null infinity 
as the generator conjugate to an asymptotic time translation. 
We derive an expression for the generator conjugate 
within the Hamiltonian framework, and propose to take this notion 
of energy as the natural generalisation of the Bondi energy 
to higher dimensions. Our definitions of asymptotic flatness 
and the Bondi energy formula differ qualitatively 
from the corresponding definitions in $d=4$; although 
the asymptotic structure of null infinity in higher dimensions  
parallels that in $4$-dimensions in some ways, the latter seems
to be a rather special case on the whole compared to general $d>4$. 
Our definitions and constructions do not work in odd spacetime dimensions, 
essentially because the unphysical metric seems to have insufficient 
regularity properties at null infinity in that case. 
} 

\section{Introduction}  \label{intro}

Gravity in higher dimensions has become one of the major 
subjects of recent studies in fundamental physics. 
There are a lot of questions that have been already  
answered in the $4$-dimensional case but remain open 
in higher-dimensions. Amongst them, perhaps the most fundamental 
issue is how to define the notion of an {\em isolated system} and 
associated {\em conserved quantities} in {\em higher-dimensions}.  
The purpose of this note\footnote{
This article is a concise version of 
the paper~\cite{HI2003}.} 
is to answer this question, providing a definition of {\em asymptotic 
flatness at null infinity in higher-dimensions}. 

\par 
In general relativity, the idea of an isolated system is given 
by defining ``asymptotic flatness'' at infinity\footnote{ 
Roughly speaking, an isolated system in general relativity is a 
spacetime that looks like Minkowski spacetime far away 
in ``any directions along spacelike or null curves.'' 
Other, less restrictive notions of an isolated system may 
also be considered, for example systems 
that look like a Kaluza-Klein space far out 
in the ``non-compact directions.'' However, the 
analysis of such metrics and of the associated conserved 
quantities would be substantially different from the ones studied here. 
}.  
One can consider two different infinities where the spacetime 
curvature vanishes: a spatial infinity and a null infinity. 
Accordingly, in $4$-dimensions, one has two notions 
of total gravitational energies: 
the ADM-energy~\cite{adm} defined at a spatial infinity and 
the Bondi-energy~\cite{bms} measured at a null infinity. 
The ADM-energy is constant. Since it is essentially 
defined by inspecting the behaviour 
of the Coulomb part of Weyl components of gravity at large spatial 
distances, it is straightforward to generalise the definition 
of the ADM-energy to the higher-dimensional case. 
On the other hand, the Bondi-energy is in general a function of time 
in the sense that it depends on the chosen cross section at null
infinity. The difference between Bondi-energies measured at two
different times represents the flux of gravitational radiation 
through the portion of null infinity bounded by the corresponding 
two cross sections.  
Since in higher-dimensions, the behaviour of the radiating part of 
gravitation is different from that of the Coulomb part, 
it is not a trivial matter to generalise the definition 
of the Bondi-energy to the higher-dimensional case. 

In order to get a sensible definition of a higher-dimensional version 
of the Bondi-energy, we first provide an appropriate generalisation 
of asymptotic flatness to higher-dimensions. 
Such a definition of asymptotic flatness should be arrived at 
by inspecting the fall off behaviour of gravitational perturbations 
near null infinity, so that the definition 
will be stable under at least linear perturbations. 
Actually, as we will see, in $d$-dimensions, perturbations typically 
drop off as $1/r^{(d-2)/2}$ as one approaches null infinity, which differs 
from the drop off rate of the Schwarzschild metric $1/r^{d-3}$ 
in higher dimensions $d>4$. Consequently our definition of asymptotic 
flatness in $d>4$ dimensions differs qualitatively from that in 
$4$-dimensions. 

We then derive an expression for the generator 
conjugate to an asymptotic time translation symmetry 
for asymptotically flat spacetimes in $d$-dimensional general
relativity ($d$ even) within the Hamiltonian framework, 
making use especially of a formalism developed by Wald and
Zoupas~\cite{wz}. This generator is given by an integral over 
a cross section at null infinity of a certain local expression 
and is taken to be the definition of the Bondi-energy in 
$d$-dimensions. Our definition yields 
a manifestly positive flux of radiated energy.  
 
\section{Asymptotic flatness at null infinity} 
\label{Asympto-flat-infinity}  

Asymptotic conditions in field theory require the specification of 
a background configuration and the precise rate at which this background 
is approached. In the case of asymptotic flatness in higher
dimensional general relativity, a natural background is 
the Minkowski metric. 
Let $(\tilde M, \tilg_{ab})$ be a $d$-dimensional physical spacetime. 
We are concerned with how to specify the precise rate 
at which Minkowski space is approached $(\tilde M, \tilg_{ab})$ 
at large distances in null directions.  

\medskip 
\noindent
{\bf Unphysical spacetime $(M, g_{ab})$: } 
It is of great technical advantage to work within a framework 
in which ``infinity'' $\I$ is attached as additional 
points to a physical spacetime manifold, $\tilde M$.  
One obtains an ``unphysical spacetime manifold'' with boundary, 
$$
M \equiv \tilde M \cup \I \,. 
$$
Furthermore, the points at infinity are brought metrically to a finite 
distance by rescaling the physical metric, $\tg_{ab}$,
by a conformal factor $\Omega^2$ with suitable properties. The asymptotic
flatness conditions are then formulated in terms of this rescaled ``unphysical
metric,''
\begin{equation}
g_{ab} \equiv \Omega^2 {\tilde g}_{ab}, 
\end{equation}
and its relation to the likewise conformally rescaled version of 
Minkowski space, to which we will refer as the ``background geometry.''

\medskip 
\noindent
{\bf Background geometry $(\bar M, \bar g_{ab})$: } 
For definiteness, we take $\bar M$ to be the region $\{-\pi/2 \le t \pm \psi \le\pi/2\}$ 
of ${\mathbb R} \times S^{d-1}$, where $t$ is the coordinate 
of $\mathbb R$ and $\psi$ is the azimuthal angle of $S^{d-1}$. 
We take the metric $\bar g_{ab}$ to be the line
element of the Einstein static universe, 
${\rm d}{s}^2=-{\rm d}t^2 + {\rm d}\psi^2 + \sin^2 \psi \,{\rm d}\sigma^2$. 
As it is well-known, the metric of the Einstein static universe is 
related to the Minkowski metric $\tilde \eta_{ab} = -{\rm d} x_0^2 +  
{\rm d} x_1^2 + \cdots + {\rm d} x_{d-1}^2$ by 
\ben
\bar g_{ab} = \Omega^2 \tilde \eta_{ab},  
\een
where $\Omega = \cos(\psi-t) \cos(\psi+t)$, 
and Minkowski spacetime corresponds precisely to the region $\bar M$.
The boundary of $\bar M$ are the conformal infinities of Minkowski 
spacetime. 

Tensor fields on Minkowski spacetime can be identified with 
tensor fields on $\bar M$, and their rate of decay at null 
infinity can be considered. To have a quantitative notion, 
we make the following definition.

\smallskip 
\noindent 
{\bf Definition: }
A tensor field, $L_{ab \dots c}$,  
is said to be {\em of order $\Omega^s$} with $s \in {\mathbb R}$, written 
$L_{ab \dots c} = O(\Omega^s)$, if the tensor field 
$\Omega^{-s} L_{ab\dots c}$ is smooth at the boundary of $\bar M$.  

\smallskip 
Consequently, if $L_{ab\dots c}$ is of order $s$, then 
$\Omega^r L_{ab\dots c}$ is of order $s+r$, and  
$\bar \nabla_{d_1} \cdots \bar \nabla_{d_k} L_{ab \dots c}$ is 
of order $s-k$. 

\subsection{Definition of asymptotic flatness} 

We now state our definition of asymptotic flatness in even 
spacetime dimensions $d>4$. 

\medskip
\noindent 
{\bf Definition } (Asymptotic flatness): 
Let $(\tilde M, \tilg_{ab})$, $(M,g_{ab})$, 
and $(\bar M, \bar g_{ab})$ be, respectively, 
a $d$-dimensional spacetime, an unphysical spacetime, 
and a background spacetime defined above, 
with smooth conformal factor $\Omega$.  
A spacetime $(\tilde M, \tg_{ab})$ is said to 
be {\it weakly asymptotically simple 
at null infinity} if the following is true: 
\begin{enumerate}
\item 
It is possible to attach a boundary, $\I$, to $\tilde M$ such that 
there exists an open neighbourhood of $\I$ in $M = \tilde M \cup \I$ 
which is diffeomorphic to an open subset of the manifold $\bar M$ of our 
background geometry, such that $\I$ gets mapped to a subset 
of the boundary of ${\bar M}$ under this identification. 
 
\item
One has, relative to our  background metric $\bar g_{ab}$, that 
\begin{eqnarray}
\label{1a}
{\bar g_{ab}} - g_{ab} 
&=& O({\Omega^{\frac{d-2}{2}}}), \quad {\bar \epsilon_{ab \dots c}} - 
\epsilon_{ab \dots c} = O({\Omega^{\frac{d}{2}}}), 
\end{eqnarray}
where $\bar \epsilon_{ab \dots c}$ and $\epsilon_{ab \dots c}$ 
denote the volume element 
(viewed as $d$-forms) associated with the metrics $\bar g_{ab}$ 
respectively $g_{ab}$, as well as
\begin{eqnarray}
\label{1b}
({\bar g^{ab}} - g^{ab}) ({\rm d} \Omega)_a 
 = O({\Omega^{\frac{d}{2}}}),
\quad ({\bar g^{ab}} - g^{ab})({\rm d} \Omega)_a({\rm d} \Omega)_b 
= O({\Omega^{\frac{d+2}{2}}}), 
\end{eqnarray}
where $g^{ab}$ is the inverse of $g_{ab}$ and where $\bar g^{ab}$ 
is the inverse\footnote{We use the convention that 
indices on tensors with a tilde are raised an lowered
with $\tilde g_{ab}$, and those with a ``bar'' are raised
and lowered with $\bar g$, and those without are with $g_{ab}$.
} of $\bar g_{ab}$.
\end{enumerate}

\medskip
\noindent
{\bf Remarks: } 
{\bf (i) } 
The physical metric remains unchanged if we change the background 
metric to $\bar g'{}_{ab} = k^2 \bar g_{ab}$ and the conformal
factor to $\Omega' = k \Omega$, with $k$ non-vanishing and smooth 
at the boundary of $\bar M$. It is easily seen that our definition 
of an asymptotically flat spacetime is independent under such a 
change of ``conformal gauge.'' 
Hence, in this sense, our definition is independent 
of the particular background geometry chosen. 

\medskip 
\noindent
{\bf (ii) }
As in $4$-spacetime dimensions, the notion of weak asymptotic simplicity 
can be strengthened by requiring in addition that every inextendible 
null geodesic in $(\tilde M, \tg_{ab})$ has precisely two endpoints 
on $\I$. Such a spacetime is then simply called {\em asymptotically 
simple}. 
This additional condition, combined with the fact that $\I$ is null, 
makes it possible to divide $\I$ into disjoint sets, $\I^+$ and
$\I^-$, on which future respectively past directed null geodesics 
have their endpoints. These sets 
are referred to as future respectively past null infinity. 
This condition also implies that $(\tilde M, \tg_{ab})$ necessarily 
has to be globally hyperbolic, by a straightforward generalisation 
of Prop.~6.9.2 of~\cite{he} to $d$-dimensions.  

\medskip 
\noindent
{\bf (iii) } 
Let us compare the above definition of asymptotic flatness with 
the behaviour of the $d$-dimensional Schwarzschild metric, 
given by the line element 
\begin{equation} 
\label{ds}
{\rm d}\tilde s^2 = - \left(1 - \frac{c}{r^{d-3}} \right) {\rm d}t^2 
+ \left(1 - \frac{c}{r^{d-3}} \right)^{-1} {\rm d}r^2 
+ r^2 {\rm d}\sigma^2, \quad c > 0. 
\end{equation}
Introducing a coordinate $u$ by the relation
${\rm d}u = {\rm d}t - (1 - cr^{-(d-3)})^{-1} {\rm d}r$, 
the line element takes the form 
\begin{equation} 
\label{ds'}
{\rm d} \tilde s^2 = -2{\rm d}u{\rm d}r 
- {\rm d}u^2 + r^2 {\rm d}\sigma^2 + cr^{-(d-3)}{\rm d}u^2.   
\end{equation} 
The first three terms on the right side are recognised 
as the Minkowski line element. 
Multiplying by our conformal factor $\Omega^2$, 
using $r^{-1} = O(\Omega)$, and using 
that $\Omega^2$ times the Minkowski metric is equal to 
our background metric ${\rm d} \bar s^2$ by construction, it follows 
that the unphysical Schwarzschild metric can be written as 
$$
{\rm d} s^2 - {\rm d} \bar s^2 =  O(\Omega^{d-1}){\rm d}u^2 
$$ 
(noting that $u$ is a good coordinate at infinity). 
It follows that Schwarzschild spacetime is asymptotically flat 
in the sense of our definition, but it becomes flat at 
null infinity at a faster rate than that specified above 
in conditions given in eqs.~\eqref{1a} and~\eqref{1b} in $d>4$. 
[In $d=4$, the relevant components drop off at the same rate, 
see the last in eqs.~\eqref{1b}.]

\medskip 
\noindent
{\bf (iv) } 
The above definition of asymptotic flatness is not 
appropriate in odd spacetime dimension, since condition~\eqref{1a} 
in item 2 now says that the unphysical metric 
$g_{ab}$ differs from the smooth background metric 
$\bar g_{ab}$ by a half odd integer power of $\Omega$, 
and thereby manifestly contradicts the assumption in item 1 
that $g_{ab}$ is smooth at the boundary. The 
powers of $\Omega$ appearing in eqs.~\eqref{1a} and~\eqref{1b} 
reflect the drop off behaviour of a linearised
perturbation (see in the next section), and it is hard to see 
how these powers could be any different from the ones  
in the full nonlinear theory. It therefore appears that 
the unphysical metric is generically at most $(d-3)/2$ times 
differentiable at the boundary in odd dimensions. We note that 
it is also inconsistent in odd dimensions to postulate that 
the quantity $\Omega^{-(d-2)/2}(g_{ab} - \bar g_{ab})$ is smooth 
at the boundary as we did above in eq.~\eqref{1a} of item 2 
in the even dimensional case, because the unphysical Schwarzschild 
metric $g_{ab}$ differs from the background 
$\bar g_{ab}$ by terms of order $\Omega^{d-1}$, 
i.e., by an even power of $\Omega$. Therefore, eq.~\eqref{1a} is
definitely false for the Schwarzschild metric in odd dimensions. 
For the Schwarzschild metric, $\Omega^{-(d-1)}(g_{ab} - \bar g_{ab})$ 
is smooth at the boundary (in even and odd dimensions), so one might 
be tempted to try this condition, together with suitable 
other conditions, as the definition of asymptotic flatness. 
However, this would eliminate from consideration all radiating 
spacetimes and is therefore not acceptable. One may try to bypass 
these problems by requiring appropriate 
lower differentiability properties of the corresponding quantities, 
but these seem neither to lead to a definition 
of asymptotic flatness that is stable under perturbations, 
nor do those weaker conditions seem to be able to guarantee 
the existence of conserved quantities such as 
the Bondi-energy.\footnote{ 
Such a difficulty in defining some quantities associated with 
radiations in odd spacetime dimensions reminds us of 
the well-known fact that in odd-dimensions the manner of radiation 
propagation is qualitatively different from that 
in even-dimensions~(see e.g.,~\cite{cdl} and references therein).  
}
Thus, it seems that a sensible definition of asymptotic simplicity 
at conformal infinity in odd spacetime dimensions would have to 
differ substantially from the one given above for even dimensions, 
and it is doubtful that such a definition can be cast into the
framework of conformal infinity.

\medskip 
\noindent 
{\bf (v) } 
We finally comment on how the above definition of asymptotic flatness 
in even spacetime dimensions $d>4$ compares to the usual definition~\cite{g} 
in $4$-dimensions. In this definition, one simply demands that there 
exists {\em some} conformal factor, $\Omega$, such that the corresponding
unphysical metric is smooth at $\I$ and such that $n_a$ is non-vanishing and 
null there. Note that the nullness of $n_a$ follows from the 
first condition if Einstein's equation 
with vanishing stress energy at null infinity are assumed.  
This definition is different in appearance from that given above 
and avoids in particular the introduction of a background geometry. 
Nevertheless, the definition of asymptotic flatness in $d=4$ 
as just stated can be brought\footnote{We emphasise, 
however, that an analogous statement is not true in $d>4$. 
Namely, it is not true that our definition of asymptotic flatness 
in higher-dimensions is equivalent to the statement that there 
exists some conformal factor, $\Omega$, such that the corresponding
unphysical metric is smooth at $\I$ and such that $({\rm d}\Omega)_a$ 
is non-vanishing and null there.} 
into a form that is very similar (but not identical) to 
the one given above for $d>4$. 
To see this in more detail, we recall that the usual definition of 
asymptotic flatness in $4$-dimensions is equivalent~\cite{tamwi} to 
the statement that with the vacuum Einstein equations, 
the physical metric can be cast into the ``Bondi form'' 
(see eqs.(14) and~(31)--(34) of \cite{bms}), 
\begin{eqnarray}
\label{bondiform}
{\rm d}\tilde s^2 &=& -2{\rm d}u{\rm d}r 
- {\rm d}u^2 + r^2 {\rm d} \sigma^2 
\nonumber\\
 &&+ O(r) {\rm d} ({\rm angles})^2 
     + O(1) {\rm d} u {\rm d} ({\rm angles}) \nonumber\\
 &&+ O(r^{-1}) {\rm d} u^2 
     + O(r^{-2}){\rm d} u {\rm d} r 
\end{eqnarray}
in suitable coordinates near null infinity, 
where the first line is recognised as the Minkowski line element
(with ${\rm d} \sigma^2$ the line element of the unit 2-sphere).

In $d>4$ spacetime dimensions
our asymptotic flatness conditions in effect state that the physical line 
element can be written in the form
\bena
\label{bondiformd}
{\rm d}\tilde s^2 &=& -2{\rm d}u{\rm d}r - {\rm d}u^2 + r^2 {\rm d} \sigma^2 \nonumber\\
           &&+ O(r^{-\frac{d-4}{2}}) {\rm d}({\rm angles})^2 
             + O(r^{-\frac{d-4}{2}}) {\rm d} u {\rm d} ({\rm angles}) \nonumber \\
           &&+ O(r^{-\frac{d-2}{2}}) {\rm d} u^2 + O(r^{-\frac{d}{2}}){\rm d} u {\rm d} r,  
\eena
where ``angles'' now stands for the polar angles of $S^{d-2}$, and ${\rm d} \sigma^2$ 
is the line element on $S^{d-2}$. 
The Bondi form~\eqref{bondiformd} in $d>4$ does not reduce to eq.~\eqref{bondiform} when $d$ is set to 4. 
The difference between the two expression arises from the ${\rm d}({\rm angles})^2$-term, which quantifies
the perturbations in the size of the cross sections of a lightcone relative to Minkowski spacetime. 
According to eq.~\eqref{bondiform}, this term is of order $O(1)$ in $d=4$ for a radiating metric, whereas
eq.~\eqref{bondiformd} would say that it ought to be of 
order $O(r^{-1})$. The latter is simply wrong for a radiating 
metric in 4 dimensions. This difference can be traced back 
to the last of conditions~\eqref{1a} in $d>4$ dimensions, 
which therefore does not hold in $d=4$. This special feature 
of $4$ dimensions will be reflected in corresponding differences 
in our discussion of the Bondi energy in dimensions $d>4$. 
We will therefore, for the rest of this paper, keep the case 
$d=4$ separate and assume throughout that $d>4$ (and even). 
Our formulas will not be valid in $d=4$ unless stated otherwise. 

\medskip
\noindent
{\bf Asymptotic symmetry: } 
A diffeomorphism $\phi$ is called an {\em asymptotic symmetry} 
if it transforms any asymptotically flat metric to an 
asymptotically flat metric. 
The asymptotic symmetries form a group under the composition 
of two diffeomorphisms. 
An infinitesimal asymptotic symmetry is a smooth vector 
field $\xi^a$ on $\tilde M$ that has a smooth extension 
(denoted by the same symbol) to the unphysical manifold, $M$, and 
which generates a 1-parameter group of asymptotic symmetries. 
It is a direct consequence of our definitions 
that the quantity 
$$\chi_{ab} \equiv \Omega^{-\frac{d-6}{2}} \pounds_{\xi} \tg_{ab}$$ 
has to satisfy 
\bena
\label{chicond}
&&\chi_{ab} = O(1), \quad \chi_a{}^a 
            = O(\Omega), \quad 
  \chi_{ab} n^a = O(\Omega), \quad 
  \chi_{ab} n^a n^b = O(\Omega^2), 
\non \\
&&{}
\xi^a n_a = O(\Omega),   
\eena
where here and in the following, $$n_a \equiv ({\rm d}\Omega)_a.$$   
Conversely, if the above relations 
are satisfied for some asymptotically flat 
spacetime, then $\xi^a$ is an infinitesimal asymptotic symmetry. 

It turns out~\cite{HI2003} that in higher dimensions, there is 
no direct analog of the infinite set of angle dependent 
translational symmetries known in $4$-dimensions.  

\medskip

One can use the freedom to choose any metric in the conformal 
equivalence class of the Einstein static universe (see (i))
to make convenient ``conformal gauge choices.'' 
A particularly useful choice for many purposes is one for which 
\begin{equation}
\label{gauge}
{\bar \nabla}_a ({\rm d} \Omega)_b = 0, \quad {\bar g^{ab}} 
({\rm d}\Omega)_a ({\rm d} \Omega)_b = 0
\end{equation}
in an open neighbourhood of the boundary, 
where $\bar g^{ab}$ is the inverse metric and $\bar \nabla_a$ 
the derivative operator associated with the background metric.
Our formula for the Bondi energy and flux assume this gauge.

\subsection{Stability of asymptotic flatness} 
We justify our definition of asymptotic flatness for even $d$ 
given above, by showing that the definition is stable under linear 
perturbations. 

\medskip
\noindent
{\bf Theorem } (Stability of asymptotic flatness):  
Let $({\tilde M},{\tilde g}_{ab})$ be a globally hyperbolic 
solution to the vacuum Einstein equations ${\tilde R}_{ab}=0$. 
Consider a solution $\delta \tilg_{ab}$ to $\delta \tilR_{ab}=0$ 
whose initial data have compact support on a Cauchy surface. Then, 
there exists a gauge such that 
setting $\delta g_{ab} = \Omega^2 \delta \tg_{ab}$,  
\bena
\label{dropoff}
&&\delta g_{ab} = O(\Omega^{\frac{d-2}{2}}), \quad 
\delta g_{ab} n^a = O(\Omega^{\frac{d}{2}}), \quad 
\non \\
&& \delta g_{ab} n^a n^b = O(\Omega^{\frac{d+2}{2}}), \quad 
g^{ab} \delta g_{ab} = O(\Omega^{\frac{d}{2}}), 
\eena 
at $\I$ for all even $d>4$. 

\medskip 

Note that these conditions are the linearised version 
of our definition of asymptotic flatness, eqs.~\eqref{1a}
and~\eqref{1b}, about an asymptotically flat background. Our definition of 
asymptotic flatness is therefore stable to linear order. 

\medskip 
\noindent
{\bf Sketch of proof: }
We choose the transverse-traceless gauge  
\bena
  {\tilde \nabla}^a \delta \tilg_{ab} 
  = \tilg^{ab}\delta \tilg_{ab} = 0 \,,  
\eena 
and define field variables by 
\bena
  \phi_\alpha = \left\{ 
        \begin{array}{@{\,}ll}
             \tau_{ab} \equiv \Omega^{-(d-2)/2} \delta g_{ab} & 
\\
             \tau_a \equiv \Omega^{-1} \tau_{ab} n^b &  
\\ 
             u \equiv \nabla^a \tau_a & 
        \end{array} 
                   \right.  \,.   
\eena 
Then we can reduce the linearised Einstein equations to the form 
\bena
  \nabla^a\nabla_a \phi_\alpha 
   = A_\alpha{}^{\beta a} \nabla_a \phi_\beta 
     + B_\alpha{}^{\beta } \phi_\beta \,, 
\label{eq:hyperbolic}    
\eena
where $A_\alpha{}^{\beta a}$ and $B_\alpha{}^{\beta }$ are smooth 
tensor fields up to and on $\I$. 
Since the hyperbolic system~(\ref{eq:hyperbolic}) possesses 
a well-defined initial value formulation~\cite{w} in the unphysical 
spacetime, we have a smooth extension of $\phi_\alpha$ on
$\I$. $\Box$ (For details, see Ref.~\cite{HI2003}.) 

\medskip 
\noindent
{\bf Remarks:} 
{\bf (i) }
In $4$-dimensions, the corresponding theorem was proved by 
Geroch and Xanthopoulos~\cite{gx}, in which 
neither the transverse-traceless gauge nor our above choice 
of the field variables work. Instead, it is necessary to 
work in the so called Geroch-Xanthopoulos gauge, 
and to take other field variables given in ref.~\cite{gx}. 
That analysis shows that the fall off 
rate of the perturbation are  
\bena
&& \delta g_{ab}= O(\Omega) \,, \quad
 \delta g_{ab}n^b= O(\Omega^2) \,, 
\non \\
&& 
 \delta g_{ab}n^an^b = O(\Omega^3) \,, \quad
 g^{ab}\delta g_{ab}= O(\Omega)   
\eena 
in $4$-dimensions.
We notice that the trace of the perturbation 
is falling off as fast as the metric 
perturbation itself. This property differs from that 
in the higher-dimensional case, where the trace falls off 
one power faster [see the last of eqs.~\eqref{dropoff}]. 

\noindent {\bf (ii) }
In odd $d$ case, since $g_{ab}$ itself is not 
smooth at $\I$, the coefficients $A_\alpha{}^{\beta a}$ 
and $B_\alpha{}^{\beta }$ are no longer smooth at $\I$, hence 
one cannot guarantee the existence of smooth solutions to the hyperbolic 
system~(\ref{eq:hyperbolic}).

\section{Gravitational energy at null infinity} 

We now define the Bondi-energy in higher-dimensions, 
using the general strategy by Wald and Zoupas~\cite{wz} for defining  
charges associated with ``boundaries''
in theories derived from a general diffeomorphism 
covariant Lagrangian. For the case of Einstein gravity,
$L = (1/16\pi G) \tilde R \tilde \epsilon$,
with the boundary given by null infinity $\I$, their scheme is
as follows. Let $\theta$ be the $(d-1)$-form 
defined by $\delta L= \mbox{Einstein's equation} + 
{\rm d}\theta$, and let $\omega$ be the symplectic current  
$\omega(\delta_1 g,\delta_2 g)=\delta_1 \theta(\delta_2 \tilg)
-\delta_2 \theta(\delta_1 \tilg)$. Further, let $\xi^a$ be 
a vector field on representing 
an infinitesimal asymptotic symmetry. If one can show that 
\begin{enumerate}
\item[({\bf 1})] 
$\omega$ has a well-defined (finite) extension 
to the $\I$ for any asymptotically flat metric, 
\item[({\bf 2})]  
there exists a symplectic potential $\Theta$ on $\I$ such that 
\bena
 ( \mbox{pullback of $\omega$ to $\I$} ) 
  = \delta_1\Theta(\delta_2 g)
     -\delta_2\Theta(\delta_1 g) \,,   
\non 
\eena 
\end{enumerate}
then the Wald-Zoupas method ensures that one can 
define an associated {\em charge} ${\cal H}_\xi$ by 
\bena 
 \delta {\cal H}_\xi =\int_B (\delta Q_\xi - \xi \cdot \theta) 
  + \int_B \xi \cdot \Theta \,, 
\label{def:charge}
\eena
where $$Q_\xi = - \frac{1}{16 \pi G} 
                  (\tilde \nabla \xi) \cdot \tilde \epsilon$$ 
is the Noether charge $(d-2)$-form, and $B$ 
a given cross section at $\I$. 

Note that it is not immediately evident that 
the above equation actually defines a charge ${\mathcal H}_\xi$
(up to an arbitrary constant), i.e., 
that the right side of eq.~(\ref{def:charge}) is indeed the ``$\delta$'' 
of some quantity. To see this, 
one first verifies that the right side of eq.~\eqref{def:charge} 
has a vanishing anti-symmetrized second variation\footnote{This 
would not be so if we had not added the $\Theta$-term to the
expression for $\delta {\mathcal H}_\xi$.}. 
This is certainly a necessary condition for it to arise as 
the  ``$\delta$'' of some quantity ${\mathcal H}_\xi$, 
for we always have $(\delta_1 \delta_2 - \delta_2 \delta_1){\mathcal
H}_\xi = 0$. 
As argued in~\cite{wz}, this is also 
a sufficient condition if one assumes that the space of asymptotically
flat metrics is simply connected. 
The arbitrary constant is fixed by setting ${\mathcal H}_\xi$ equal 
to zero on Minkowski spacetime.

One can show that the assumptions ${\bf (1)}$ and 
${\bf (2)}$ that are needed for the existence of $\cal H_\xi$ 
indeed hold under our choice of the boundary conditions 
(fall off conditions) for asymptotic flatness. Namely,  
one can show that the pullback of $\omega$ to $\I$ can be written 
in terms of the smooth variables $\tau_{ab}$, $\tau_a$ at $\I$, 
and that $\Theta$ can be given in terms of a smooth tensor 
field at $\I$ with vanishing trace, called the  

\smallskip \noindent 
{\bf News tensor in higher-dimensions: }  
$$
 N_{ab} \equiv \mbox{pullback to $\I$ of} \,\, 
 \Omega^{-(d-4)/2}q^m{}_a q^n{}_b S_{mn} \,, 
$$
where $q_{ab} \equiv g_{ab}-2n_{(a}l_{b)}$, 
with $l_al^a=0,\,\,n^al_a=+1$, is the projection onto $\I$ 
and $S_{ab}$ is defined by $(d-1)(d-2)S_{ab} 
\equiv 2(d-1) R_{ab} - Rg_{ab}$. 
In fact, $\Theta$ is expressed as\footnote{ 
Although $\Theta$ has a ``gauge freedom'' with respect to 
the change of the conformal factor $\Omega$, this conformal gauge 
freedom can actually be fixed by imposing the gauge condition~\eqref{gauge} 
on the background metric $\bar{g}_{ab}$~\cite{HI2003}, 
which is seen to yield the following conditions on the unphysical metric,  
$$
  n^an_a= O(\Omega^{(d+2)/2}),\quad 
  \nabla_a n_b = O(\Omega^{(d-2)/2}) . 
$$
Our results for the Bondi energy formula are obtained 
under this gauge fixing.  
} 
$ \Theta_{a_1 \dots a_{d-1}} = (1/32\pi G)  
\tau^{cd} N_{cd} \, \epsilon_{a_1 \dots a_{d-1}}   
$. 


\medskip 
If $\xi^a$ is {\em asymptotic time translation}, then it can 
be written as 
$\xi^a = \alpha n^a - \Omega \nabla_a \alpha$ with $\alpha$ a suitable
function that specifies the translation in question.
We now restrict our consideration to the special case of such 
translation asymptotic symmetries.
Then, as the explicit expression of ${\cal H}_\xi$, we obtain  


\smallskip 
\noindent
{\bf Bondi-energy formula in even dimensions $d>4$: } 
For any such infinitesimal translation, the Bondi energy
(-momentum) is given by
\begin{multline}
 {\cal H}_\xi 
 = \frac{1}{8(d-3)\pi G} \int_B
   \Omega^{-(d-4)} 
   \bigg( 
         \frac{1}{(d-2)} R_{ab} q^{ac}  q^{bd} (\nabla_c l_d)\xi^e l^f\\ 
        -\Omega^{-1} \alpha^{-1} (l^{[e} -v \nabla^{[e}\log \alpha)
         C^{f]bcd} \xi_b (l_c - v \nabla_c \log \alpha) \xi_d 
   \bigg) \, \epsilon_{efa_1 \cdots a_{d-2}} \,, 
\label{bondimass}
\end{multline}
where $v$ is  
defined by $\nabla_a v = l_a$, and ${\epsilon}_{efa_1 \cdots a_{d-2}}$ 
denotes the natural volume element on $(M,g_{ab})$, and where 
we are assuming that $\alpha$ is such that $\xi^a$ corresponds to a null-translation
to keep the formula simple. 

We also have the flux formula associated with such a $\xi^a$ 
through a segment $\S$ of $\I$ 
 \begin{equation}
 \label{fluxf2}
 F_{\xi} = - \frac{1}{32\pi G} \int_\S 
\alpha N^{cd} N_{cd} \, {^{(d-1)} \epsilon}. 
\end{equation}
For positive $\alpha$, $\xi^a$ is future directed timelike or null 
at null-infinity.
This shows that the flux of energy (defined via {\em any} asymptotic 
future-directed translational symmetry) through $\I$ is always negative, 
i.e., that the energy radiated away by the system is always positive. 

\medskip 

In the case $d=4$, the energy formula~\eqref{bondimass} 
is not correct and needs to be modified by replacing $1/(d-2) R_{ab}$ 
by the combination $(1/2)(S_{ab} - \rho_{ab})$, see~\cite{wz}. 
It then coincides with an expression for the quantities associated 
with asymptotic translations first proposed by Geroch~\cite{g}. 

The first and second term in the integrand of~\eqref{bondimass} 
can be roughly interpreted as follows: the second term is 
the ``Coulomb part'' of the Weyl tensor (multiplied by suitable 
powers of $\Omega$), and represents the ``pure Coulomb contribution'' 
to the Bondi energy. 
The first term represents contributions from gravitational radiation; 
it follows from the conditions for the vector field $l^a$ that 
it vanishes if and only if 
the news tensor, $N_{ab}$, and hence the flux, vanishes. 
In $4$-dimensions, it can be proved~\cite{g} that the news tensor, 
and hence the radiative contribution to the Bondi energy, 
always vanishes in stationary spacetimes. 
It would be interesting to see whether an analog 
of this result holds in $d$-dimensions. 

\medskip
In the $d$-dimensional Schwarzschild spacetime,  
the Bondi energy is evaluated as follows. 
The term involving $R_{ab}$ in our expression~\eqref{bondimass} 
for the Bondi energy does not contribute, showing that there is 
no radiative contribution to the Bondi energy. 
The Coulomb contribution is found to be 
$\Omega^{-(d-3)} C^{abcd} l_a n_b l_c n_d = c(d-2)(d-3)/4$ at $\I$. 
Normalising $\xi^a$ so that it coincides with  
the timelike Killing field $t^a$ of the metric~\eqref{ds}. 
Inserting this into our Bondi-energy formula, we get 
\begin{equation}
\mathcal H_{\alpha n} 
= \frac{c(d-2)A_{d-2}}{16 \pi G} \quad \quad \text{($= \frac{c}{2G}$ 
in $4$-dimensions).} 
\end{equation} 
This coincides with the ADM mass of the spacetime~\eqref{ds} 
(given e.g. in~\cite{mp}), as we expect.  

\medskip 
The Bondi-energy at a cross section $B \subset \I$ 
is interpreted as, naively  speaking, the ADM-energy minus 
the energy of gravitational radiation emitted by some isolated system 
in the causal past of $B$. If the Bondi-energy of an isolated system 
became negative, that would imply that the system radiates away 
more energy than it had initially. It is not possible to tell 
from the above integral expression if the Bondi-energy is positive or not, 
and therefore a further analysis is needed. In the $4$-dimensional 
case, positivity was confirmed in~\cite{LV,SY,HP}, but for 
higher-dimensions, this issue is still open.

\end{document}